\documentclass[twocolumn,showpacs,preprintnumbers,nofootinbib,prd,superscriptaddress,groupedaddress,10pt]{revtex4-1}

\usepackage{graphicx,amssymb,amsmath,amsthm,amsfonts,epsfig,epsf}
\usepackage[linktocpage]{hyperref}
\usepackage[usenames]{color}
\usepackage{epstopdf}

\usepackage{aas_macros}
\usepackage{bm}
\usepackage{dcolumn}
\usepackage[latin1]{inputenc}
\usepackage{latexsym}
\usepackage{rotating}
\usepackage{hyperref}
\usepackage{color}
\usepackage{longtable}
\usepackage{enumerate}
\usepackage{tensor}
\usepackage{url}
\def\nn{\nonumber}

\newcommand{\ben}{\begin{enumerate}}
\newcommand{\een}{\end{enumerate}}

\def\be{\begin{equation}}
\def\ee{\end{equation}}
\def\bea{\begin{eqnarray}}
\def\eea{\end{eqnarray}}
\newcommand{\beq}{\begin{eqnarray}}
\newcommand{\eeq}{\end{eqnarray}} 
\newcommand{\ba}{\begin{align}}
\newcommand{\ea}{\end{align}}

\def\nn{\nonumber}

\begin{document}

\title{Accretion of dark matter by stars 
%Structure and growth of dark-hearted stars
}

\author{
Richard Brito$^{1}$,
Vitor Cardoso$^{1,2}$ %\footnote{Electronic address: vitor.cardoso@tecnico.ulisboa.pt},
Hirotada Okawa$^{3,4}$
}
\affiliation{${^1}$ CENTRA, Departamento de F\'{\i}sica, Instituto Superior T\'ecnico -- IST, Universidade de Lisboa -- UL,
Avenida Rovisco Pais 1, 1049 Lisboa, Portugal}
\affiliation{${^2}$ Perimeter Institute for Theoretical Physics Waterloo, Ontario N2J 2W9, Canada}
\affiliation{${^3}$ Yukawa Institute for Theoretical Physics, Kyoto University, Kyoto, 606-8502, Japan}
\affiliation{${^4}$ Advanced Research Institute for Science \& Engineering,
Waseda University, 3-4-1 Okubo, Shinjuku, Tokyo 169-8555, Japan}

\begin{abstract}
Searches for dark matter imprints are one of the most active areas of current research.
We focus here on light fields with mass $m_B$, such as axions and axion-like candidates.
Using perturbative techniques and full-blown nonlinear Numerical Relativity methods, we show that 
(i) dark matter can pile up in the center of stars, leading to configurations and geometries oscillating with frequency which is a multiple of $f=2.5\times 10^{14}\,\left(m_{B}c^2/eV\right)\,{\rm Hz}$.
These configurations are stable throughout most of the parameter space, and arise out of credible mechanisms for dark-matter capture.
Stars with bosonic cores may also develop in other theories with effective mass couplings, such
as (massless) scalar-tensor theories. We also show that (ii) collapse of the host star to a black hole is avoided by efficient gravitational cooling mechanisms.
\end{abstract}

%\tableofcontents
%\end{widetext}
%\clearpage

%95.35.+d 	Dark matter (stellar, interstellar, galactic, and cosmological) (see also 95.30.Cq Elementary particle processes; for brown dwarfs, see 97.20.Vs; for galactic halos, see 98.35.Gi or 98.62.Gq; for models of the early %Universe, see 98.80.Cq)
%14.80.-j 	Other particles (including hypothetical)
%11.10.St 	Bound and unstable states; Bethe-Salpeter equations
%12.60.-i 	Models beyond the standard model (for unified field theories, see 12.10.-g)
%04.25.D-    Numerical relativity
%04.25.dc    Numerical studies of critical behavior, singularities, and cosmic censorship
%04.25.dg    Numerical studies of black holes and black-hole binaries
%04.25.-g    general relativity: approximation methods, equations of motion
%04.40.-b 	Self-gravitating systems; continuous media and classical fields in curved spacetime
%04.50.-h    Higher-dimensional gravity and other theories of gravity
%04.50.Cd    KaluzaKlein theories
%04.50.Gh    Higher-dimensional black holes, black strings, and related objects
%04.60.Cf    Gravitational aspects of string theory
%04.70.-s    Physics of black holes
%04.70.Bw    Classical black holes
%04.70.Dy    Quantum aspects of black holes, evaporation, thermodynamics
%04.80.-y    Experimental studies of gravity
%04.80.Cc    Experimental tests of gravitational theories
%11.25.Mj    Compactification and four-dimensional models
%11.10.Kk    Field theories in dimensions other than four

\maketitle

%\tableofcontents
%%%%%%%%%%%%%%%%%%%%%%%%%%%%%%%%%%%%%%%%%%%%%%%%%%%%%%%%%%%%%%%%%%%%%%%%%%%%%%
\noindent{\bf{\em I. Introduction.}}
%%%%%%%%%%%%%%%%%%%%%%%%%%%%%%%%%%%%%%%%%%%%%%%%%%%%%%%%%%%%%%%%%%%%%%%%%%%%%%
The standard picture for the evolution and structure of our universe relies
on the existence of unseen forms of matter, generically called dark matter (DM).
The evidence for DM in observations is overwhelming, starting with galaxy rotation curves,
gravitational lensing and the cosmic microwave background.
While carefully concocted modified theories of gravity can perhaps explain almost all observations,
the most attractive and accepted explanation lies in DM being composed mostly of cold, collisionless particles.

Ultralight fields, such as axions or axion-like candidates are an attractive possibility~\cite{Marsh:2015wka,2014NatPh..10..496S,2014PhRvL.113z1302S}. Axions were originally devised to solve the strong-CP problem, but recently a plethora of other, even lighter fields with masses $10^{-10}-10^{-33}\,{\rm eV}/c^2$, have also become an interesting possibility, in what is commonly known as the axiverse scenario~\cite{Arvanitaki:2009fg}. The simplest possible theory is that of a massive scalar $\phi$ or vector $A_{\mu}$ minimally coupled to gravity, and described by the Lagrangian
\be
{\cal L}=\frac{R}{\kappa} - \frac{F^2}{4}- \frac{\mu_V^2}{2}A_{\nu}A^{\nu}-\frac{g^{\mu\nu}}{2}\phi^{\ast}_{,\mu}\phi^{}_{,\nu} - \frac{\mu_S^2}{2}\phi^{\ast}\phi\,.\label{eq:MFaction}
\ee
We take $\kappa=16\pi$, $F_{\mu\nu} \equiv \nabla_{\mu}A_{\nu} - \nabla_{\nu} A_{\mu}$ is the Maxwell tensor and $F\equiv F^{\mu\nu}F_{\mu\nu}$. The mass $m_B$ of the boson under consideration is related to the mass parameter above through $\mu_{S,V}=m_{B}/\hbar$, and the theory is controlled by the dimensionless coupling
\begin{equation}
\frac{G}{c\hbar} M_T\mu_{S,\,V} = 7.5\cdot 10^{9} \left(\frac{M_T}{M_{\odot}}\right) \left(\frac{m_{B}c^2}{eV}\right)\,,
\end{equation}
where $M_T$ is the total mass of the bosonic configuration.

It is appropriate to emphasize that DM has not been seen nor detected through any of the known standard model interactions.
The only evidence for DM is through its gravitational effect. Not surprisingly, the quest for DM is one of the most active fields of research of this century.
Because DM interacts feebly with Standard Model particles, and thanks to the equivalence principle, the most promising channel to look for DM imprints consists of gravitational interactions.
\begin{figure}[htb]
\begin{center}
\begin{tabular}{c}
\epsfig{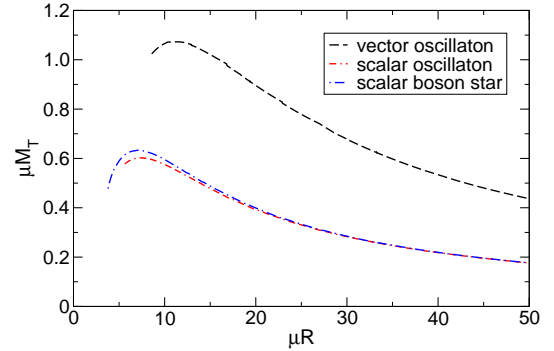}
\end{tabular}
\caption{Comparison between the total mass of a boson star ({\it complex} scalar) and an oscillaton ({\it real} scalar or vector fields), as a function of their radius $R$. $R$ is defined as the radius containing 98\% of the total mass. The procedure to find the diagram is outlined in the main text.\label{MvsR}}
\end{center}
\end{figure}
It turns out that {\it classical} massive bosonic fields minimally coupled to gravity, described by Eq.~\eqref{eq:MFaction} -- which will be our working model for DM here -- can form structures~\cite{Kaup:1968zz,Ruffini:1969qy,Seidel:1991zh,Comment:2015,Guth:2014hsa}. Self-gravitating {\it complex} scalars may give rise to static, spherically-symmetric geometries called boson stars, while the field itself oscillates~~\cite{Kaup:1968zz,Ruffini:1969qy} (for reviews, see Refs.~\cite{Jetzer:1991jr,Schunck:2003kk,Liebling:2012fv,Macedo:2013jja}). On the other hand, {\it real} scalars have a non-trivial time-dependent stress-tensor but may give rise to long-term stable oscillating geometries called oscillatons~\cite{Seidel:1991zh}. Both solutions arise naturally as the end-state of gravitational collapse~\cite{Seidel:1991zh,Garfinkle:2003jf,Okawa:2014nea}, and both structures share a similar mass-radius relation, summarized in Fig.~\ref{MvsR}. In this figure we also show the mass-radius relation for massive vectors, which as far as we know has not been discussed before. Boson stars and oscillatons have a maximum mass $M_{\rm max}$, given approximately by
\be
\frac{M_{\rm max}}{M_{\odot}}=8\times 10^{-11}\,\left(\frac{\rm eV}{m_{B}c^2}\right)\,,\label{max_mass}
\ee
for scalars and slightly larger for vectors. Oscillatons are not truly periodic solutions of the field equations, as they decay through quantum and classical processes. However, their lifetime $T_{\rm decay}$ is extremely large for the masses of interest~\cite{Page:2003rd,Grandclement:2011wz},
\be
T_{\rm decay}\sim 10^{324}\left(\frac{1\,{\rm meV}}{m_B{\rm c^2}}\right)^{11}\, {\rm yr}\,.
\ee
%

%%%%%%%%%%%%%%%%%%%%%%%%%%%%%%%%%%%%%%%%%%%%%%%%%%%%%%%%%%%%%%%%%%%%%%%%%%%%%%
\noindent{\bf{\em II. Compact stars with DM cores.}}
%%%%%%%%%%%%%%%%%%%%%%%%%%%%%%%%%%%%%%%%%%%%%%%%%%%%%%%%%%%%%%%%%%%%%%%%%%%%%%
Although not framed in the context of DM capture, some stars with complex scalar fields in their interior were considered before in the literature~\cite{Henriques:1989ar,Henriques:1989ez,Lopes:1992np,Henriques:2003yr,Sakamoto:1998aj,deSousa:2000eq,deSousa:1995ye,Pisano:1995yk}. Our construction is more generic.

Consider the Lagrangian \eqref{eq:MFaction}, augmented with the stress-energy tensor for a perfect fluid,
\be
T_{\rm fluid}^{\mu\nu}=\left(\rho_{F}+P\right)u^{\mu}u^{\nu}+P g^{\mu\nu}\,,
\ee
with $u^{\mu}$ the fluid four-velocity, $u^{\mu}=\left(\sqrt{-1/g_{tt}},0,0,0\right)$, and $\rho_F,\,P$ its density and pressure.
It is easy to show that the vector field must satisfy the constraint $\mu^2_V\nabla_{\mu}A^{\mu}=0$, while the Bianchi identities and the Klein-Gordon equation, impose the conservation equations $\nabla_{\mu}T_{\rm fluid}^{\mu\nu}=0$. Let us focus for simplicity on scalar fields. For complex scalars, self-gravitating solutions were considered many times, consisting of spherically symmetric, {\it static} boson stars~\cite{Jetzer:1991jr,Schunck:2003kk,Liebling:2012fv,Macedo:2013jja}. In Ref.~\cite{Seidel:1991zh} it was shown that \emph{real} massive scalars admit oscillating
self-gravitating solutions. Consider therefore a general time-dependent, spherically symmetric geometry
\be\label{metric}
ds^2=-F(t,r)dt^2+B(t,r)dr^2+r^2d\Omega^2\,.
\ee
Rescaling the scalar field as $\phi\to \phi/\sqrt{8\pi}$ and defining the function $C(t,r)=B(t,r)/F(t,r)$, the field equations then lead to the system of partial differential equations
\beq
&&B'/B=(r/2)\left[C\dot{\phi}^2+(\phi')^2+B\left(\mu_S^2\phi^2+16\pi\rho_{F}\right)\right]\nn\\
&&+(1-B)/r\,,\label{cont_eq1}\\
%
%\dot{B}/B&=&r\dot{\phi}\phi'\,,\qquad 2\dot{\rho_{F}}= r\left(P+\rho_{F} \right)\dot{\phi}\phi'\,,\label{cont_eq1}\\
%
&&C'/C=2/r+Br\left(\mu_S^2\phi^2+8\pi\rho_{F}-8\pi P\right)-2B/r\,,\\
&&\ddot{\phi}C=-\dot{C}\dot{\phi}/2+\phi''+2\phi'/r-C'\phi'/(2C)-B\mu_S^2\phi\,,\\
&&2P'=-\left(P+\rho_{F}\right)\left(CB'-BC'\right)/(BC)\,.\label{cont_eq2}
\eeq
These equations suggest the following periodic expansion
\beq
N^i&=&\sum_{j=0}^{\infty} N^i_{2j}(r)\,\cos\left(2j\omega t\right)\,,\nonumber\\
\phi(t,r)&=&\sum_{j=0}^{\infty} \phi_{2j+1}(r)\,\cos\left[\left(2j+1\right)\omega t\right]\,,\label{series_fluid}
\eeq
where $N^i=(B(t,r),C(t,r),\rho_{F}(t,r),P(t,r))$.
The equations of motion need to be supplemented by an equation of state. We will focus exclusively, and for simplicity, on a polytropic equation of state $P=K \rho_{F}^{\Gamma}$ with $K=100$ and $\Gamma=2$, which can mimic neutron stars~\cite{1964ApJ...140..434T}, but our results generalize to other equations of state~\footnote{In geometrical units $G=c=1$ and $\Gamma=2$, $K$ has units length$^2$ and can be used to set the length-scale of the problem in the absence of the scalar field~\cite{1964ApJ...140..434T}. The choice $K=100$ was used in e.g. Ref.~\cite{ValdezAlvarado:2012xc}.}. 
Inserting the expansion \eqref{series_fluid} into the system~\eqref{cont_eq1}--~\eqref{cont_eq2} and truncating the series at a given $j$~\footnote{We find that the solutions typically converge already for $j=2$ for most of the parameter space.}, gives a set of ordinary differential equations for the radial Fourier components of the metric functions and the scalar field. These equations need to be solved imposing regularity at the origin and asymptotic flatness.

A useful quantity to describe scalar-fluid stars is the scalar field's energy density, given by
\be\label{scalar_density}
2\rho_{\phi}=-2T_0^{\phantom{0}0}=\dot{\phi}^{2}/g_{tt}+\phi^{'2}/g_{rr}+\mu_S^2\phi^2\,.
\ee
%
%where $T^{\mu\nu}$ is the scalar field's stress energy tensor. 
With this, we define the time-average total mass in the fluid and bosons as
\be
M_{F,\,B}=\int_0^{\infty} 4\pi \left<\sqrt{B}\,\rho_{F,\,B}\right> r^2 dr\,,\label{fermion_mass}
\ee
where $<>$ denotes a temporal average. The total mass $M_T$ can be found in the usual way through the metric component $g_{rr}$ which asymptotically approaches the Schwarzschild solution at infinity.  
The procedure to find fluid-vector stars is similar and will not be described further. Figure~\ref{MvsR} shows a phase diagram of solutions when there is no fluid. We find that the maximum mass for a stable scalar oscillaton is $M_T\sim 0.6/\mu_S$ in agreement with previous studies~\cite{Seidel:1991zh,Alcubierre:2003sx}. 
As far as we know, the vector solution has not been described anywhere else.

\begin{figure}[htb]
\begin{center}
\begin{tabular}{ccc}
\epsfig{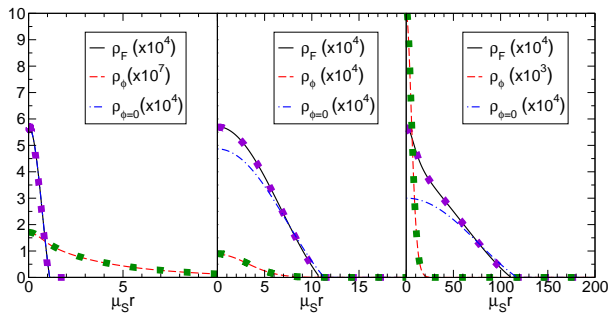}%\epsfig{file=./plots/scalarfluid_rhovsrhoscal_mu01_MBperc21.eps,width=5.1cm,angle=0,clip=true}&
\end{tabular}
\caption{Comparison between the (time average) energy density of the scalar field $\rho_{\phi}$ and the fluid $\rho_{F}$ for mixed scalar oscillatons and fermion fluids.
From left to right, $\mu_S M_0=0.1,\,M_B/M_T\approx 21\%$, % corresponding to $\phi_1(0)=0.026$, $\rho_{F\,0}(0)=0.00057092$ and $\omega M_0\approx 0.0993$.
%
%Middle: 
$\mu_S M_0=1,\,M_B/M_T\approx 5\%$,% corresponding to $\phi_1(0)=0.064$, $\rho_{F\,0}(0)=0.00057092$ and $\omega M_0\approx 0.873$.
%
%Right: 
and $\mu_S M_0=10,\,M_B/M_T\approx 5\%$. %, corresponding to $\phi_1(0)=0.06982$, $\rho_{F\,0}(0)=0.00057092$ and $\omega M_0\approx 8.629$.
Squares denote the corresponding quantities for complex fields (i.e. mixed boson-fluid stars). The overlap is nearly complete.
Here $M_0$ and $\rho_{\phi=0}$ are the total mass of the star and the energy density of the fluid, respectively, when the scalar field vanishes everywhere. In the left panel, the $\rho_{\phi=0}$ and the $\rho_{F}$ lines are indistinguishable, because light fields have a negligible influence on the fluid distribution.
\label{scalarfluid_osci}}
\end{center}
\end{figure}

When the fluid is present, the solutions depend on two parameters $\rho_{F\,0}(0)$ and $\phi_1(0)$, and can be characterized by the mass coupling $\mu_S M_0$, where $M_0$
is the mass of the \emph{static} star for vanishing scalar field, corresponding to the same value of central density $\rho_{F\,0}(0)$. Different configurations are shown in Fig.~\ref{scalarfluid_osci},
where we compare the scalar energy density against the fluid's density. In all cases we also show the density profile when the scalar field is trivial, corresponding to the same fluid mass $M_F$.
The overall behavior might have been anticipated from an analysis of Fig.~\ref{MvsR}: for light fields, $\mu_S M_0<1$, the scalar profile is extended and the pure oscillaton solution
is broad and light. As such, the scalar has a negligible influence on the fluid distribution (as can be seen from the fact that the zero-scalar line $\rho_{\phi=0}$ overlaps with the fluid line),
and these stars simply have an extended scalar condensate protruding away from them.
\begin{figure}[htb]
\begin{center}
\begin{tabular}{c}
\epsfig{file=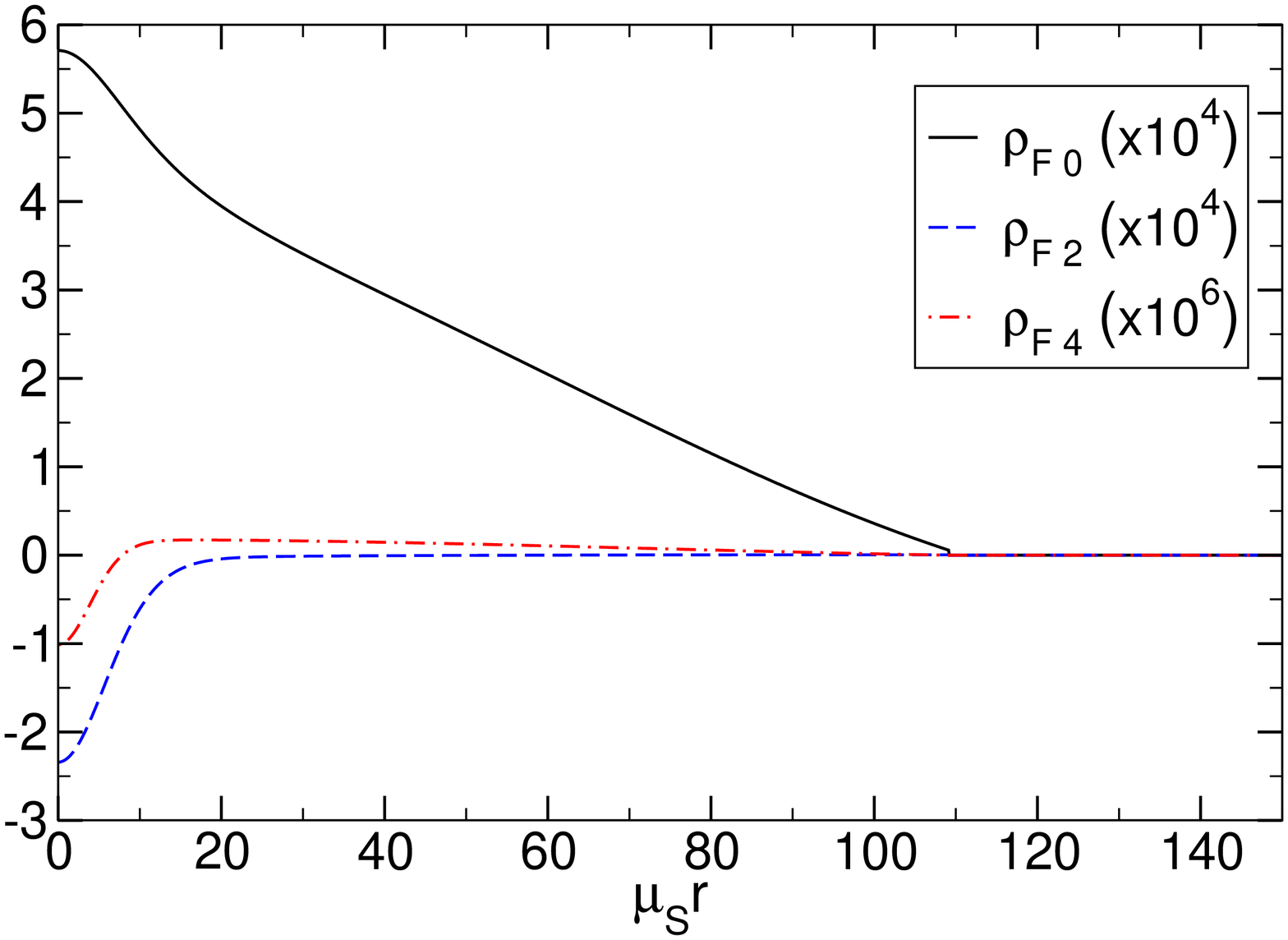,width=6.9cm,angle=0,clip=true}
\end{tabular}
\caption{Density profile of the fluid at $t=0$ and its first Fourier components for $\mu_S M_0=10$ and $M_B/M_T\approx 5\%$.
\label{scalarfluid_osci2}}
\end{center}
\end{figure}
In fact, our results are compatible with a decoupling between the boson and fluid for large $\mu_S M_B$. For this case Fig.~\ref{MvsR} alone is enough to interpret the bosonic distribution.
For example, for $\mu_S M_0=10, M_B/M_T=5\%$, we get $\mu_S M_B\sim 0.3$, which would imply from Fig.~\ref{MvsR} that $\mu_S R\sim 20$ for the scalar field distribution. This is indeed apparent from Fig.~\ref{scalarfluid_osci}.
Similar conclusions were reached when studying mixed fermion fluid/boson stars with complex fields~\cite{Lopes:1992np}~\footnote{In fact, the structure of mixed oscillatons and fluid stars is almost identical to that of geons and fluids, as can be seen from Fig.~\ref{scalarfluid_osci}, where we overplot with dotted lines the complex field case.}.

%%%%%%%%%%%%%%%%%%%%%%%%%%%%%%%%%%%%%%%%%%%%%%%%%%%%%%%%%%%%%%%%%%%%%%%%%%%%%%
\noindent{\bf {\em Global thermalization.}}
%%%%%%%%%%%%%%%%%%%%%%%%%%%%%%%%%%%%%%%%%%%%%%%%%%%%%%%%%%%%%%%%%%%%%%%%%%%%%%
We should stress that the composite stars we study here in general {\it oscillate}, driven by the scalar field.
For these configurations, the convergence of the series~\eqref{series_fluid} is rather fast, meaning that the amplitude of the oscillations is relatively small, but nonzero. For high scalar field central densities, first-order terms $j=1$ in the density might become of the order of the zeroth-order term, making it difficult to find these configurations with good accuracy. For a given $\rho_{F\,0}(0)$, we expect these configurations to become unstable at some threshold $\phi_1(0)>\phi^c_{1}(0)$. Our results are consistent with what was previously found for boson-fermion fluid stars~\cite{Henriques:1989ar,Henriques:1989ez}. Although field configurations with high $M_B/M_T$ are more challenging to find for large $\mu_S M_0$, we expect them to follow the same kind of behavior as that found in boson-fermion stars.
An example showing the amplitude of the oscillations is shown in Fig.~\ref{scalarfluid_osci2} for $M_0\mu_S=10$, for which the star is oscillating with large amplitudes (the oscillating component is of the same order of magnitude as the static one). We find that even for $M_B/M_T=0.01$, and for $\mu_S M_0=10$ the oscillations are of order $10\%$ of the static component.
These oscillations imply that both the scalar field and the fluid density (which is coupled to it gravitationally), vary periodically with a frequency~\footnote{The fundamental frequency $\omega\sim \mu_{S,\,V}$~\cite{Seidel:1991zh,Page:2003rd}.}
\be
f=2.5\times 10^{14}\,\left(\frac{m_{B}c^2}{eV}\right)\,{\rm Hz}\,,
\ee
or multiples thereof. For axion-like particles with masses $\sim 10^{-5} \,eV/c^2$, these stars would emit in the microwave band.
These oscillations are driven by the boson core and might have observable consequences; it is in principle even possible that resonances occur when the frequency of the scalar is equal to the oscillation frequency of the unperturbed star. 
The joint oscillation of the fluid and the boson might be called a {\it global thermalization of the star}, and is expected to occur also
for boson-star-like cores (which give rise to static boson cores), once the scalar is allowed to have non-zero couplings with the star material. 

We have neglected viscosity in the star's fluid and local thermalization. Viscous timescales for neutron star oscillations can be shown to be large compared to the star dynamical timescale $R$,
but small when compared to the (inverse of) the accretion rate likely to be found in any realistic configuration~\cite{1990ApJ...363..603C}. As such, we expect that viscosity will damp global oscillations of the star, eventually leading to a depletion of the scalar field core. A similar effect will occur with local thermalization
of the scalar with the star material; detailed studies of these effects are still necessary.

\begin{figure*}[ht]
 \begin{tabular}{ccc}
  \psfig{file=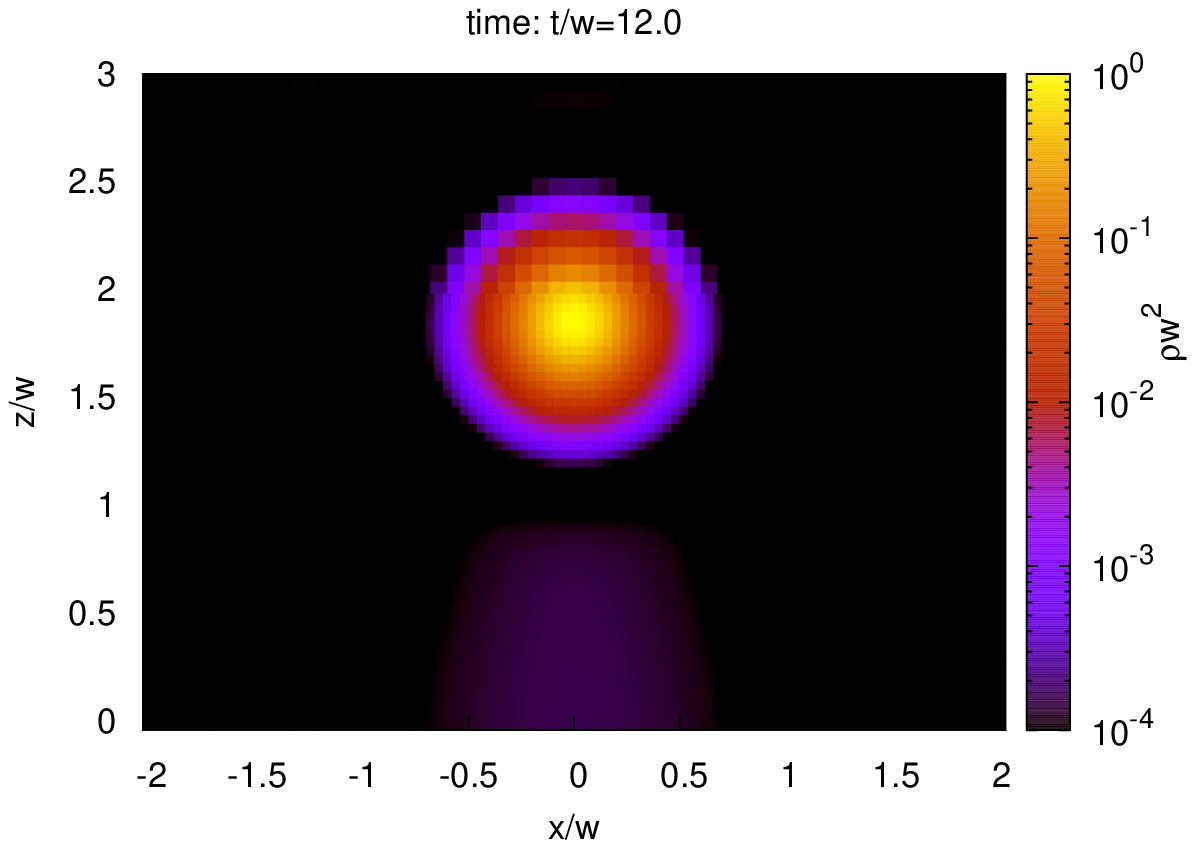,width=5.7cm}\hspace{1em}%&
  \psfig{file=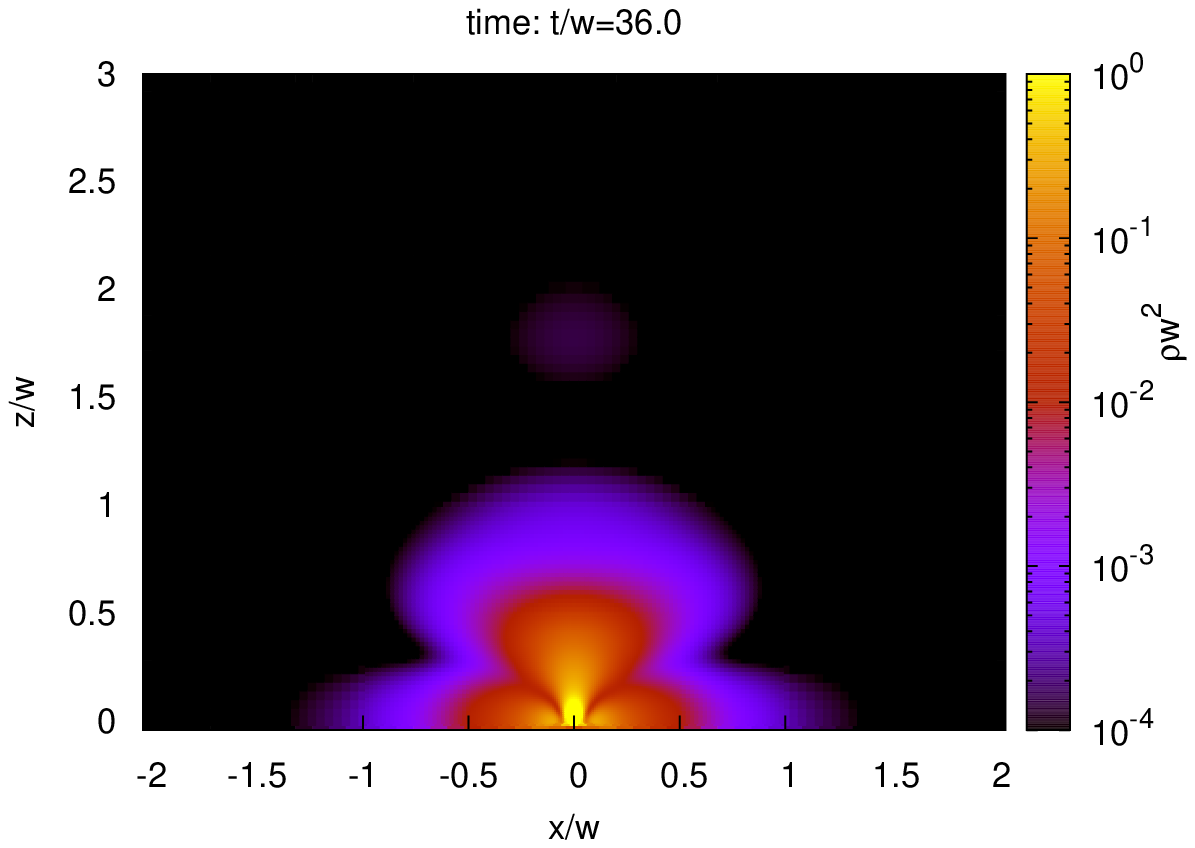,width=5.7cm}\hspace{1em}%&
  \psfig{file=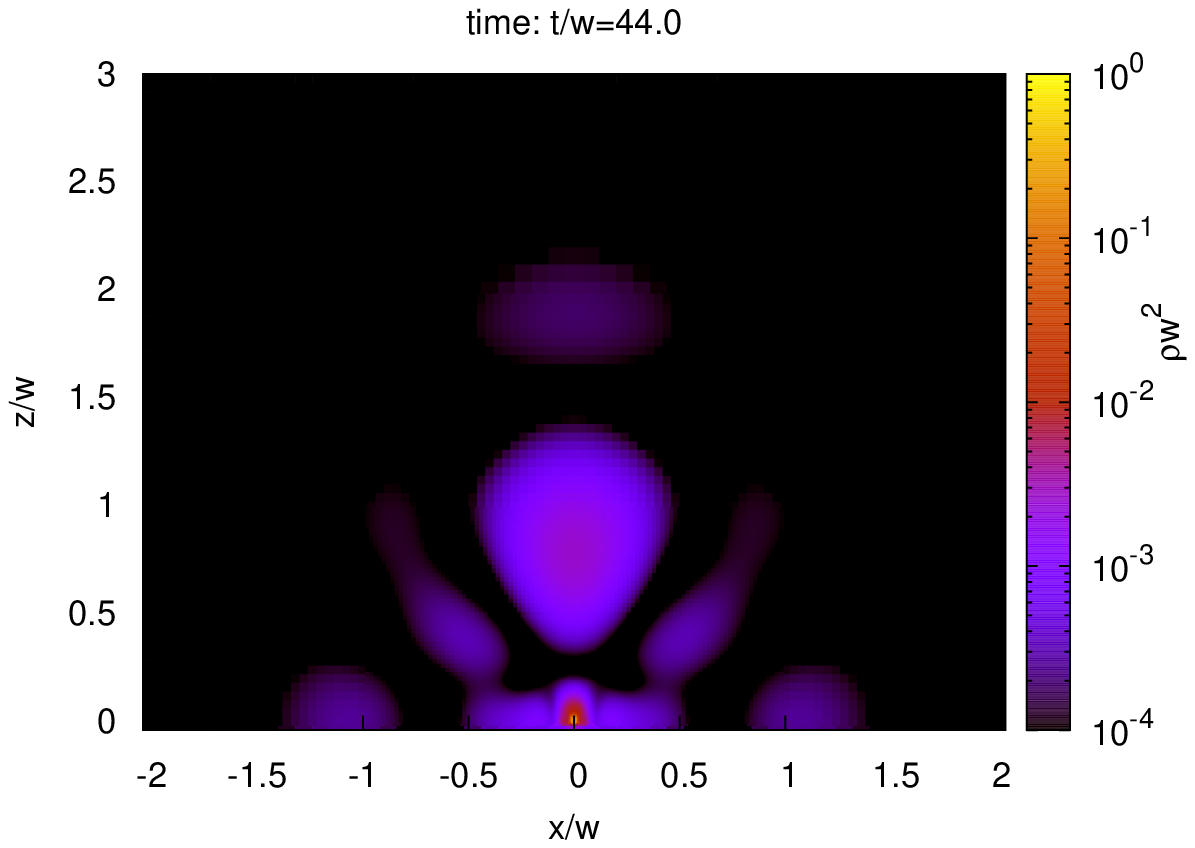,width=5.7cm}
 \end{tabular}
 \caption[]{Snapshot of density~$\rho$ during the collision of two equal-mass oscillatons, each with $M\mu_S\sim 0.5, R\mu_S\sim 15$, along the plane of collision, for half the space (the remaining space can be obtained by symmetry). 
 Notice how the final configuration is more compact, and massive. At very late times, we find that it relaxes to a point on the curve of Fig.~\ref{MvsR}.
 }
 \label{fig:res_col_A1.2}
\end{figure*}
%
%%%%%%%%%%%%%%%%%%%%%%%%%%%%%%%%%%%%%%%%%%%%%%%%%%%%%%%%%%%%%%%%%%%%%%%%%%%%%%
\noindent{\bf{\em Stability.}}
%%%%%%%%%%%%%%%%%%%%%%%%%%%%%%%%%%%%%%%%%%%%%%%%%%%%%%%%%%%%%%%%%%%%%%%%%%%%%%
Finally, the stability of these solutions cannot easily be inferred from a mass versus radius diagram, and it requires a dynamical analysis which goes beyond the scope of this Letter. 
However, one can argue that a {\it necessary} condition for stability is that the binding energy $M_T-M_F-M_B$ be negative~\cite{Henriques:1989ez}. Although negative values do not necessarily imply stability, they do give strong support to the claim that these configurations are stable. A more careful stability analysis 
shows that for sufficiently small $\phi_1(0)$ and for stars which are stable in the absence of scalars, the negativity of the binding energy is a good criterion for stability~\cite{Henriques:1990xg}. 
To summarize, for sufficiently small scalars composite stars are dynamically stable. On the other hand, our results show that these configurations can be understood well from the mass-radius relation of oscillatons.
The maximum mass supported is (\ref{max_mass}), $M_{\rm max}/M_{\odot}=8\times 10^{-11}\,eV/(m_{B}c^2)$, which for a neutron star and an axion field of mass $10^{-5}\,eV$ falls well within the stability regime~\cite{Gleiser:1988}.

%%%%%%%%%%%%%%%%%%%%%%%%%%%%%%%%%%%%%%%%%%%%%%%%%%%%%%%%%%%%%%%%%%%%%%%%%%%%%%%%%%%%%%%%%%%%%%%%%%
\noindent{\bf{\em III. Formation: DM accretion, core growth and gravitational cooling mechanism.}}
%%%%%%%%%%%%%%%%%%%%%%%%%%%%%%%%%%%%%%%%%%%%%%%%%%%%%%%%%%%%%%%%%%%%%%%%%%%%%%%%%%%%%%%%%%%%%%%%%%
We have shown that pulsating stars with DM cores exist as solutions of the field equations. Do they form dynamically?
Pulsating purely bosonic states certainly do~\cite{Seidel:1991zh,Garfinkle:2003jf,Okawa:2014nea}. There are two different channels
for formation of composite fluid/boson stars. One is through gravitational collapse in a bosonic environment, through which the star is born already with a DM core. The second process consists of capture and accretion of DM into the core of compact stars~\cite{Press:1985ug,Gould:1989gw}.

Once the scalar is captured it will interact with the boson core. Interactions between complex fields have shown that
equal mass collisions at low energies form a bound configuration~\cite{Bernal:2006ci,Palenzuela:2006wp}. In other words, two bosonic cores composed of complex fields interact and form a more massive core at the center. This new bound configuration is in general asymmetric and will decay on large timescales~\cite{Macedo:2013jja}, the final state being spherical~\cite{Bernal:2006it}. 
We have repeated this analysis for two oscillatons colliding at sufficiently small energies, using the methods of Numerical Relativity outlined in Refs.~\cite{Okawa:2014nda,Okawa:2014nea}. We find that generically the collision of two oscillatons whose total mass is below the peak value will give rise to a more compact, more massive oscillaton configuration. By contrast, the collision of oscillatons whose mass exceeds the peak value results in large mass losses and a lighter final configuration.
An example of such a collision is shown in Fig.~\ref{fig:res_col_A1.2} for equal-mass oscillatons each with $M_T\mu_S\sim 0.5$ (close to the peak value of Fig.~\ref{MvsR}). Note that the total mass is larger than the peak value
and one would naively predict gravitational collapse to a black hole (BH). Instead, the collision first produces a massive, more compact oscillating object, accompanied by large scalar field losses, as is apparent from Fig.~\ref{fig:res_col_A1.2}. On large timescales the configuration relaxes to a very low-mass structure.

This is a general feature of what has been termed the ``gravitational cooling mechanism:'' a very efficient (dissipationless) mechanism that stops them from growing past the unstable point, through the ejection of mass~\cite{Alcubierre:2003sx,Seidel:1993zk}. Such features have been observed in the past in other setups, such as spherically symmetric gravitational collapse
(see Fig. 2 in Ref.~\cite{Okawa:2014nea}), or slightly perturbed oscillatons~\cite{Alcubierre:2003sx}. Gravitational cooling provides a counter-example to an often-used {\it assumption} in the literature: that stars accreting DM will grow past the Chandrasekhar limit for the DM core and will collapse to a BH~\cite{Gould:1989gw,Bertone:2007ae,McDermott:2011jp,Graham:2015apa,Kouvaris:2013kra}. Our results show that this need not be the case, if the DM core is prevented from growing by a self-regulatory mechanism, such as gravitational cooling.

Thus, even though other more detailed simulations are still needed, the likely scenario for evolution would comprise a core growth through minor mergers, slowing down close to the mass-radius peak (see Fig.~\ref{MvsR}), at which point it stops absorbing any extra bosons~\cite{Alcubierre:2003sx,Okawa:2014nea}. In other words, the unstable branch is never reached. This phenomenology is specially interesting, as it would also provide
a capture mechanism for these fields which is independent of any putative nucleon-axion interaction cross section:
as we discussed, the bosonic core grows (in mass) through accretion until its peak value. At its maximum, it has a size $R_B/M_{\odot}\sim M_B/M_{\odot}$. This is the bosonic core {\it minimum size}, as described by Fig.~\ref{MvsR}. Even for $M_B=0.01M_T$ the core has a non-negligible size and is able to capture and trap other low-energy oscillatons.

%%%%%%%%%%%%%%%%%%%%%%%%%%%%%%%%%%%%%%%%%%%%%%%%%%%%%%%%%%%%%%%%%%%%%%%%%%%%%%
\noindent{\bf{\em IV. Conclusions.}}
%%%%%%%%%%%%%%%%%%%%%%%%%%%%%%%%%%%%%%%%%%%%%%%%%%%%%%%%%%%%%%%%%%%%%%%%%%%%%%
Previous works on the subject of DM accretion by stars have implicitly assumed that
the DM core is able to grow without bound and eventually collapse to a BH~\cite{Gould:1989gw,Bertone:2007ae,McDermott:2011jp,Graham:2015apa,Kouvaris:2013kra}.
Our results, from full nonlinear simulations of the field equations, show that the core may stop growing when it reaches a peak value, at the threshold
of stability, if DM is composed of light massive fields. Gravitational cooling quenches the core growth for massive cores and the core growth halts, close to the peak value (c.f. Fig.~\ref{MvsR}).

Finally, our results are formally valid only for the theory~\eqref{eq:MFaction}, and require the fields to be massive. Minimally coupled, multiple (real) scalars, interacting only gravitationally, were also shown to give rise to similar configurations~\cite{Hawley:2002zn}. We have explicitly verified that similar results may hold in other setups. For example, scalar-tensor theories (which can be shown to give rise to an effective position-dependent mass term~\cite{Cardoso:2013fwa,Cardoso:2011xi}) and non-minimally coupled {\it massless scalars} are, in principle, also able to develop nontrivial
configurations with a scalarized core. Further details will be presented in a forthcoming work.

%\clearpage
%\newpage
%%%%%%%%%%%%%%%%%%%%%%%%%%%%%%%%%%%%%%%%%%%%%%%%%%%%%%%%%%%%%%%%%%%%%%%%%%%%%%
\noindent{\bf{\em Acknowledgments.}}
%%%%%%%%%%%%%%%%%%%%%%%%%%%%%%%%%%%%%%%%%%%%%%%%%%%%%%%%%%%%%%%%%%%%%%%%%%%%%%
%\begin{acknowledgments}
%
We thank E. Berti, L. Gualtieri, I. Lopes, D. Marsh and U. Sperhake for useful comments and feedback.
R.B. acknowledges financial support from the FCT-IDPASC program through the grant SFRH/BD/52047/2012, and from the Funda\c c\~ao Calouste Gulbenkian through the Programa Gulbenkian de Est\' imulo \`a Investiga\c c\~ao Cient\'ifica.
V.C. acknowledges financial support provided under the European
Union's FP7 ERC Starting Grant ``The dynamics of black holes: testing
the limits of Einstein's theory'' grant agreement no. DyBHo--256667,
and H2020 ERC Consolidator Grant ``Matter and strong-field gravity: New frontiers in Einstein's theory'' grant agreement no. MaGRaTh--646597. Research at Perimeter Institute is supported by
the Government of Canada through Industry Canada and by the Province
of Ontario through the Ministry of Economic Development $\&$
Innovation.
We acknowledge allocations on SDSC Trestles and Comet and TACC Stampede through NSF-XSEDE Grant PHY-090003.
%
%\end{acknowledgments}
%%%%%%%%%%%%%%%%%%%%%%%%%%%%%%%%%%%%%%%%%%%%%%%%%%%%%%%%%%%%%%%%%%%%%%%%%%%%%%
%\vskip 5mm

\bibliographystyle{h-physrev4}
\bibliography{Ref}

\end{document}